\documentstyle{article}

\def\be{\begin{eqnarray}}
\def\ee{\end{eqnarray}}
\def\nn{\nonumber}

\hoffset= - 1.0in         

\begin{document}

\hfill ITEP/TH-15/08

\bigskip

\centerline{\Large{Alday-Maldacena duality
and AdS Plateau problem
}}

\bigskip

\centerline{A.Morozov}

\bigskip

\centerline{\it ITEP, Moscow, Russia}

\bigskip

\centerline{ABSTRACT}

\bigskip

{\footnotesize
A short summary of approximate approach to the
study of minimal surfaces in AdS, based on
solving Nambu-Goto equations iteratively.
Today, after partial denunciation of the BDS conjecture,
this looks like the only constructive approach
to understanding the ways of its possible modification
and thus to saving the Alday-Maldacena duality.
Numerous open technical problems
are explicitly formulated throughout the text.
}

\bigskip

\centerline{\it Contribution to Proceedings of Osaka workshop,
OCU, December 2007}

\bigskip

\bigskip

\section{Plateau problem}

Plateau problem \cite{Wass}, named after the blind Belgian physicist
who originated the systhematic study of complicated configurations
of soap bubbles, is one of the most long-standing
and poorly understood problems in modern mathematics,
despite enormous attention and effort.
It is about explicit construction in $d$-dimensional
space of a $2d$ surface $S$
with a given boundary $\partial S = \Pi$
which has an extremal area, i.e.
Nambu-Goto functional $A(S)$:
\be
{\rm find}\ S: \ \ \ \ \ \left.
\frac{\delta A}{\delta S} = 0\ \right|_{\ \partial S = \Pi}
\ee
As usual for a prominent mathematical problem it is
simultaneously a very important one in {\it string theory},
where its solutions provide  classical approximations
to string amplitudes and thus to the Yang-Mills
amplitudes in the strong coupling phase \cite{SGD,GM}.
Therefore any idea,  allowing to predict the
strong-coupling asymptotics of Yang-Mills amplitudes
from perturbative field-theory considerations,
should immediately provide a big step towards solution
of Plateau problem.
To some this intimate connection to one of the big
"unsolvable" problems is a serious additional argument
against any such ideas,
nicknamed weak-strong or simply $S$-dualities,
while for the others it can be a source of additional
enthusiasm.
An important "detail" is that in the stringy context
the minimal surface is embedded not into the original
space-time of the Yang-Mills theory, but into a space
with additional Liouville dimension and sophisticated
(often unknown) metric.

The metric is believed to be known and simple
in the case of ${\cal N}=4$ super Yang-Mills theory:
the relevant space-time is $AdS_5$ \cite{AdS/CFT}
(additional $S_5$
factor is inessential for Plateau problem considerations).
In the same theory the situation with amplitudes is
also simplified, and reasonable conjectures are already made
about their behavior in strong-coupling regime.
Inspired by the old Sudakov summation of leading logarithms
\cite{LLA},
these strong-coupling continuations (SCC)
assume one-or-another kind of
exponentiation of perturbative amplitudes \cite{ABDK},
i.e. existence of enormously strong relations between higher-
and lower-loop diagrams, and look like ordinary Wilson
averages, but in auxiliary (momentum!) space.
This average is just abelian under the overoptimistic BDS
hypothesis \cite{BDS}
(i.e. all higher loops are expressed through the
one-loop expressions), and some SCC of non-abelian average
in more realistic, but less ambitious -- and less informative --
conjectures.
The hypothetical identity between AdS minimal areas
and BDS-style Wilson averages is 
known as {\it Alday-Maldacena duality} \cite{AM1}.

\section{The present status of Alday-Maldacena duality
\label{presta}}

Alday-Maldacena duality \cite{AM1}-\cite{IMM2}
was originally motivated by a ground-breaking
BDS hypothesis \cite{BDS} and stated that
\cite{AM1,DKS1,BHT,MMT1}:
\be
\exp (A_\Pi) = \exp\Big(\frac{\kappa}{R^2}\times
{\rm minimal\ area}\Big)
\ \stackrel{?}{=}\ {\bf abelian}\ {\rm Wilson\ loop\ average}\ \
= \exp (D_\Pi)
\nn \\
{\rm where} \ \ \
\exp (D_\Pi) = \left< \exp \oint_\Pi A_\mu(y) dy^\mu \right>
= \exp \left(
\oint_\Pi\!\oint \frac{dy^\mu dy'_\mu}{(y-y')^2}
\right)\ \ \ \ \ \ \ \ \ \
\label{AMD-BDS}
\ee
which one can call a BDS version of Alday-Maldacena duality.
If this was true, this could actually imply explicit
resolvability of Plateau minimal-surface problem,
at least in the AdS geometry
(reservation: strictly speaking the surface itself
could still be not constructed explicitly,
only its area needs to be known).
This would be very interesting, but somewhat suspicious.

Today BDS hypothesis is known to be slightly(?) violated
\cite{B-V} and current modification of (\ref{AMD-BDS})
is considerably weaker \cite{DHKS01,DHKS03}:
\be
\exp\Big(\frac{\kappa}{R^2}\times
{\rm minimal\ area}\Big) \stackrel{?}{=}
SCC\Big({\bf non-abelian}\ {\rm Wilson\ loop\ average}\Big)
\label{AMD-DHKS}
\ee
One can call this statement
DHKS version of Alday-Maldacena duality,
and it is far less radical from the point of view of
Plateau problem: the r.h.s.
of (\ref{AMD-DHKS}) is also quite transcendental,
actually, double transcendetal:
first, because of sophisticated structure of non-abelian
Wilson average, second, because of an absolutely mysterious
SCC operation (whose role was reduced to renormalization
of the coupling constant in the case of (\ref{AMD-BDS})).

\bigskip

Direct study/check of Alday-Maldacena duality involves
several separate problems.

Difficult problem: find minimal surface and its regularized
area. Usually this can be done only approximately:
as expansion in some small parameters for some special
kinds of the boundary shapes (small parameters characterize
the shape deviation from some solvable configurations \cite{IMM2}
or are just numerically small \cite{IMM1,IM8}).

Simple problem: find abelian Wilson averages.

Somewhat difficult, but straightforward problem:
find the lowest terms in weak coupling expansion of
non-abelian Wilson average.

Esoteric (mysterious) problem: extract strong-coupling non-abelian
Wilson average from these first terms of weak-coupling
expansion.
Alternatively one can try to extract weak-coupling
expansion from {\it approximate} formulas for the minimal
area: this problem is equally mysterious.

In this sense the relation (\ref{AMD-DHKS}) is not very different
from the general belief of the string-gauge duality \cite{SGD}:
it is a probable but not too constructive statement.
In fact, (\ref{AMD-DHKS}) is even more puzzling,
because there is no clear physical reason for representation
of multi-loop Yang-Mills amplitudes in the form of
a Wilson average in momentum space, even in perturbative regime,
but quite a lot of numerical evidence which supports this.

\bigskip

In the absence of physically clear motivation for
(\ref{AMD-DHKS}) one naturally leans to the search of
high-science explanations: for example, one can think that
the both sides of (\ref{AMD-DHKS}) belong to the same
universality class of generalized $\tau$-functions
\cite{gentau}, controlled by some hidden integrability
structure.
A little closer-to-the-Earth version of such approach
would be an assumption that
some explicit set of Virasoro-like constraints (loop
equations) can be written down for both sides of the
equation.
This last hope is slightly supported by discovery
of a certain conformal symmetry, respected by both
sides \cite{DKS1,Koma}: this is a finite-dimensional
global group, resembling the $SL(2)$ subalgebra of
Virasoro algebra, which is {\it not}, however, immediately
continued to entire infinite-dimensional half of the Virasoro
algebra, at least naively \cite{IMM2}.

\bigskip

From a radical point of view (\ref{AMD-DHKS}) is not
any strong-weak duality at all, at least at the moment:
too small is known about its r.h.s.
Somewhat more optimistic is an assumption that
(\ref{AMD-BDS}) is violated only "slightly" and can be
somehow explicitly corrected.
The deviations from (\ref{AMD-BDS}) discovered in \cite{AM3} and
\cite{B-V} are numerically small, while those found in \cite{IMM2}
are, perhaps, not small numerically, but instead they
affect only very few of infinitely many structures appearing
at the two sides of the duality relation.
Moreover, the approach of \cite{IMM2} seems to provide some
way to {\it calculate} the deviation from (\ref{AMD-BDS}):
the problem is to recognize the possible structures behind
the formulas.
In the case of success this can actually provide a clue to
the correct {\it definition} of the r.h.s. of (\ref{AMD-DHKS}).

\bigskip

In what follows we give a sketchy review of the most important
results from \cite{IMM1,IM8,IMM2} and speculate on their
possible use in the future research of AdS Plateau problem
and Alday-Maldacena duality.

\section{Solving Plateau problem approximately}

The key suggestion of \cite{IMM1,IM8,IMM2} is to look for
"perturbative" solution to Plateau problem,
expanding it in some small parameter(s) for the boundary
$\Pi$, slightly deviating from a few exactly solvable cases.

Then the sophisticated non-linear Euler-Lagrange equations,
implied by Nambu-Goto (NG) action (area functional)\footnote{
We use notation $y^\mu$ for coordinates of embedding space
(actually, $AdS_5$) in order to emphasize that this is
auxiliary (in fact, momentum \cite{GM,AM1}) space and not the
physical space-time.
The minimal surface and its boundary $\Pi$
-- composed from the $n$ null-momenta of scattering gluons --
lies in this space.
Coordinates on the $2d$ world sheet of
a string are denoted $z$ or $\zeta$.}
\be
A(S) = \int_S \sqrt{\det_{a,b=1,2}
G_{\mu\nu}(y)\frac{\partial y^\mu}{\partial z^a}
\frac{\partial y^\nu}{\partial z^b}}\ d^2\!z
\ee
are substituted by iterative sequence of linear equations
\be
\Delta_{NG}^{(G)} y^\mu_{m+1} =
{\cal O}_m^{\,\mu} (y_0,\ldots,y_m|G)
\label{recuproc}
\ee
where the r.h.s. is a non-linear combination
of lower iterations, and $\Delta_{NG}$ at the l.h.s.
is peculiar Nambu-Goto differential operator, which
depends on the metric in embedding space.
In the case of $AdS_5$ space
\be
\Delta_{NG}^{AdS} = \Delta_0 - {\cal D}^2 + {\cal D}
= 4 \partial\bar\partial - \bar z^2\bar\partial^2 -
2z\bar z\partial\bar\partial - z^2\partial^2
\ee
where $\Delta_0 = 4\partial\bar\partial$ and
${\cal D} = z^a\frac{\partial}{\partial z^a} =
z\partial + \bar z\bar\partial$
are the flat Laplace and dilatation operators respectively.
The dilatation operator part is divided by $AdS$ radius
which is put equal to one everywhere in this paper, $R=1$.

Thus the problem of iterative solution of
Plateau problem is reduced to:

$\bullet$ inverting $\Delta_{NG}$ operator,

$\bullet$ accounting for the boundary conditions,

$\bullet$ summing up iteration series $\sum_k y_k$,

$\bullet$ regularization of the action/area integral.

Of course, the most interesting phenomena in Plateau theory
are related to possible divergencies of the iteration
series, i.e. to the third item in the list:
this is what stands behind appearance of multiple
and singular solution and this is what the abstract theory
of Plateau problem \cite{Wass} is largely focused on.
However, for the study of Alday-Maldacena duality this looks
like inessential details: the problems appear much earlier,
when solutions are still unique and smooth and no problems with
series divergency are expected.
Thus at the present stage of development
essential are the other three of above problems.

\section{The theory of $\Delta_{NG}$ operator}

Ideally one should (and can) develop this theory to the
same extent as that of its flat Laplace analogue.
Considerable complication is that its zero modes --
{\it NG harmonic functions} -- are not just combinations
of holomorphic and anti-holomorphic functions, instead
they are equally simple but far less familiar Legendre
functions of negative semi-integer index $-3/2$,
members of respectable hypergeometric family.
They were iteratively constructed in \cite{IM8} and
finally expressed as linear combinations of {\it elementary}
functions in \cite{IMM2}:
\be
\Delta_{NG}\left(\sum_{k=0}^\infty \big(u_k z^k + v_k\bar z^k\big)
\frac{\Big(1+k\sqrt{1-z\bar z}\Big)
\Big(1-\sqrt{1-z\bar z}\Big)^k}{(z\bar z\big)^k}
\right) = 0, \ \ \ \ \forall u_k, v_k
\label{NGharf}
\ee
Of course, all $z\bar z$ are divided by $AdS$ radius $R=1$,
they would disappear, giving rise to the ordinary
harmonic functions, in the flat limit $R=\infty$.

To perform iteration procedure (\ref{recuproc}) one needs
to work out explicit formulas for ${\cal O}_k^{AdS}$
-- what is tedious but straightforward exercise (which still
remains to be done, see \cite{IMM2} for the first non-trivial
expressions) -- and to resolve the equations.
The latter step involves construction of a Green function
for $\Delta_{NG}$, what at the moment remains an open problem.
Solving the equations is a little less straightforward
than writing them down, because there are ambiguities in the choice
of the zero modes (\ref{NGharf}), which are actually controlled by
the boundary conditions.

\section{Boundary conditions and the boundary rings}

As explained and illustrated in detail in \cite{IMM1,IM8},
appropriate imposition of boundary conditions is a delicate
task in the study of Plateau problem in the context of
Alday-Maldacena duality.
The reason for this is the divergency of the area integral,
which (divergency) comes from the region near the boundary $\Pi$
and predominantly near its {\it corners}.
Because of this we are interested in {\it approximate}
solution which is {\it exact} in the vicinity of the boundary
and especially accurate at places where the boundary is not smooth.
This is not quite a standard requirement for approximate
solutions and one needs to develop some dedicated methods
to handle the problem.
In particular, corners can be effectively controlled by the
methods of non-linear algebra \cite{nolal}.

Approach suggested and partly developed in \cite{IMM1,IM8}
is to use the {\it boundary ring} ${\cal R}_\Pi$ of the boundary,
requiring to look for solutions to (\ref{recuproc}) not
among all possible functions, but only among restricted set,
belonging to the ring and thus satisfying the boundary
conditions {\it a priori}.
Immediate new problem is that expansion (\ref{NGharf})
is not in very good accordance with the boundary ring
structure: the individual terms in (\ref{NGharf}) are
not elements of the ring.
Actually this means that for every given boundary one
should substitute (\ref{NGharf}) by a new expansion with basis
chosen consistently: made from the elements of the
corresponding ${\cal R}_\Pi$.
In \cite{IMM1,IM8} the iterative construction of
such basises is described not only for rotational symmetric
boundary (\ref{NGharf}), but also for some polygonial shapes
(including generic admissible rectangulars of \cite{MMT1}
-- an exhaustive generalization of original $n=4$ solution of
\cite{Kru,AM1}),
but it is not yet developed to the same level of explicitness
as was (\ref{NGharf}) in \cite{IMM2}.

Even without a fully developed theory the boundary ring
approach provides excellent approximations to
minimal surfaces, and it would be extremely interesting to
perform numerical comparison of the corresponding area
with the r.h.s. of (\ref{AMD-BDS}) and, perhaps, with some
anticipated corrections implied by its substitution with
(\ref{AMD-DHKS}).
The latter comparison is of course a matter of art,
because of the lack of a clear definition for SCC operation.
However, even before this kind of interesting duality problems
can be addressed, one still needs to convert approximate
minimal surfaces into equally good approximate
expressions for their {\it regularized} areas.

\section{Regularization and enhanced regularization dependence}

Evaluation of regularized areas is a separate and quite
a sophisticated issue.
The reason for divergency is that the boundary $\Pi$ is
located at the absolute of $AdS$ space, at $r=0$, where
$AdS$ metric $ds^2 = \frac{dy_0^2 - d\vec y^2 - dr^2}{r^2}$
is singular.
One can think of different ways of regularization.
One possibility is to change the metric, say, write
$r^2+\mu^2$ instead of $r^2$ in denominator
-- we call this $\mu$-regularization in \cite{IMM2}.
Its drawback is that it actually changes the theory,
moreover explicitly breaks conformal invariance, which
is the global symmetry of original problem.
This in turn violates various Ward identities and one
should be careful about potential anomalies.
Another possibility is to preserve the theory, but
change the observable, say, shift $\Pi$ by a small
distance $c$ away from the absolute.
This option, which we call $c$-regularization,
still suffers from various ambiguities.
First of all, there is no {\it canonical} way to make such shift:
there is no distinguished prescription for what is $\Pi_c$
after the shift.
Second, $\Pi$  plays two {\it a priori} different roles:
it is the place where the boundary conditions are imposed
and it also restricts the domain of integration in $A(S)$.
In fact, there is no need to attribute these two roles
to the same $\Pi$: as soon as there is also a shifted $\Pi_c$,
whatever it is, one can, for example, use this $\Pi_c$
as a boundary of integration domain, while keep
boundary conditions imposed at original $\Pi$.
From our experience with renormalizable theories we
usually neglect all these possible complications, because
the answer there is universal: whatever of many different
choices we make, the answer does not change, up to just
a few regularization-dependent parameters.
This, however, does not need to be true in the non-local context
of Alday-Maldacena duality, and in \cite{IMM2} it is explicitly
shown that regularization details do matter, at least in
the strong coupling regime and at least for $n=\infty$.
Theoretical understanding of this phenomenon remains to be
found, but, importantly enough, even in this $n=\infty$
case some restricted class of regularizations provide the
stable result for regularized area, thus making it
regularization-{\it in}dependent, at least in this restricted
sense \cite{IMM2}.
Actually, $\mu$-regularization belongs to this class,
and it was also shown to work nicely in the $n=4$ case,
originally considered in \cite{AM1,MMT1}.
More care is needed if $c$-regularization is used,
explicit prescriptions can be formulated to make results
consistent with $\mu$-regularization, but careful theoretical
examination is still lacking.

Additional aspect, which should be kept in mind in consideration
of regularized actions, concerns interplay between
the Nambu-Goto and $\sigma$-model formulations of the
minimal-action problem.
The $\sigma$-model action is considerably simpler than the NG
one, at the same time it is well known that NG solutions
can be reproduced from the $\sigma$-model solutions if
additional Virasoro constraint is imposed \cite{MMT2,Yang}.
This makes substitution of Nambu-Goto problem by the
its $\sigma$-model analogue seemingly attractive and some
early calculations, including \cite{AM1} and \cite{MMT1}
were made in this formalism.
There are, however, two problems with this "simplified"
approach.
First, solving $\sigma$-model equations can be actually
more difficult than NG ones \cite{IMM1},
because one should additionally fix coordinate dependence in
coordinate-independent NG solution.
Second, regularization can break the equivalence between
NG and $\sigma$-model actions and it actually {\it does},
even in the simplest $n=4$ example \cite{Popo}.
In this example emerging difference is just a constant
and is not physically significant, however, it can easily
become more important for higher values of $n$.

\section{$n=\infty$ case: a wavy circle}

The most advanced results about approximate minimal surfaces
and their regularized areas
are so far obtained in \cite{IMM2}, in the academic case
of scattering of $n=\infty$ ultra-soft gluons, when the
boundary of minimal surface is an arbitrary smooth curve.
Among exactly solvable are two important cases:
of two parallel lines and of a circle.
The shape of AdS minimal surface between two parallel lines is
explicitly described by elliptic integral \cite{parali}
and this explicit formula was used in \cite{AM3}
to establish the first violation of (\ref{AMD-BDS}).
However, there are practically no free parameters in this
example and it is much more interesting to study the
deviations of $\Pi$ from the straight lines.
This problem of "wavy lines" was first addressed in
\cite{PoRy,Sem}, but it was long before the era of Alday-Maldacena
duality and the problem was not analyzed so ambitiously,
therefore today it needs to be re-addressed and re-examined.
We started such re-examination in \cite{IMM2},
but from another exactly solvable example: the one of a circle.
This choice has two obvious technical advantages.
First, the shape of $AdS$ minimal surface with a circle boundary
at the absolute $r=0$ is described by an elementary function:
$\vec y^{\,2} + r^2 = 1$, there is no need for anything like
elliptic integrals.
Second, a slightly perturbed circle has positive external
curvature everywhere, while for a wavy line which is perturbation
of a straight line, curvature is frequently changing sign.
For minimal surface problem this is a real technical complication,
because in the vicinity of the boundary the surface leans
towards the curvature center and its projection onto the plane
$r=0$ lies on one side of the boundary in the case of the
wavy circle, while it jumps from one side of the wavy
straight line to the other when the boundary curvature changes sign,
and this complicates parametrization of the surface.

In fact, even the choice of appropriate parametrization of the
wavy line is a technically important issue.
If we restrict consideration to $\Pi$ which are {\it planar}
deformations of the circle, i.e. lie entirely inside this
circle's plane, then complex coordinates on this plane
can be conveniently used.
In particular, according to Riemann theorem,
any curve $\Pi$ can be parameterized by a complex-analytic
function:
\be
\Pi:\ \ \ \ z = \zeta + \sum_{k=0}^\infty h_k\zeta^k,
\ \ \ \ \zeta = e^{i\phi}
\label{holpar}
\ee
This parametrization has many advantages, but also
is not free from drawbacks.
The main of them is sophisticated realization of conformal
and probably the other relevant symmetries, which can
act by mixing $z$ and $r$ coordinates -- and for minimal
surface $r$ is a non-holomorphic function, depending
on both $z$ and $\bar z$.
Since Riemann theorem is far from providing a constructive
transformation to the form (\ref{holpar}),
it is a separate problem to evaluate the action
of such symmetries on parameters $h_k$.

The first terms of $A(S)$ expansion in powers of $h_k$
were found in \cite{IMM2}:
\be
\begin{array}{ccccccccccccccc}
{A_\Pi} & =& \frac{\pi}{2\mu}L_\Pi &-& 2\pi &
-& 3\pi\Big(Q^{(2)}_\Pi &-&
Q^{(3)}_\Pi &-& 3Q^{(3,diag)}_\Pi &+& O(h^4)\Big) &&
\label{Aans}
\end{array}
\ee
Divergent term is proportional to the length of the
boundary curve
\be
L_\Pi = 2\pi
\sqrt{1+h_1}\sqrt{1+\bar h_1}\left(1 + \frac{1}{4}\sum_{k=2} {k^2
|h_k|^2\over \left|1+h_1\right|^2} - \ \ \ \ \ \ \ \ \ \ \ \ \
\right. \nn \\ \left. -
\frac{1}{16}\sum_{k,l=2}^\infty
kl(k+l-1) \left[{h_kh_l\bar h_{k+l-1}\over
\left(1+h_1\right)^2\left(1+\bar h_1\right)} + {\bar h_k\bar h_l
h_{k+l-1}\over \left(1+\bar h_1\right)\left(1+\bar h_1\right)^2}
\right]+ O(h^4)\right),
\ee
while the most interesting finite terms are made from the
following structures:
\be
Q^{(2)}_\Pi = \sum_{k=0}^\infty B_k|h_k|^2, \ \ \ \ \ \ \ \
B_k = \frac{k(k-1)(k-2)}{6},
\ee
\vspace{-0.3cm}
\be
Q^{(3)}_\Pi = 
\frac{1}{2}\sum_{i,j=0}^\infty C_{ij}\Big(h_ih_j\bar h_{i+j-1} +
\bar h_i\bar h_j h_{i+j-1}\Big), \ \ \ \ \  
C_{ij} = \frac{ij}{6}\Big(i^2+3ij+j^2-6i-6j+7\Big)
\ee
with diagonal part of $Q^{(3)}_\Pi$,
\be
Q^{(3,diag)}_\Pi = \frac{1}{2}\sum_{i=0}^\infty C_{ii}
\Big(h_i^2\bar h_{2i-1} + \bar h_i^2 h_{2i-1}\Big),
\ \ \ \ \ \ \ \ \ \
C_{ii} = \frac{i^2}{6}\Big(5i^2-12i+7\Big) =
\frac{i^2(i-1)(5i-7)}{6},
\ee
providing a separate additional contribution
(it is also contributing through the $Q^{(3)}_\Pi$ term,
so the total coefficient in front of $Q^{(3,diag)}_\Pi$
in brackets is $1+3=4$).

The calculation of area in this order is relatively simple,
because one needs to know solution of NG equations
with only $h$-linear terms included -- this is because
we are expanding in the vicinity of exact solution, --
and the $h$-linear terms in NG equations
are actually $\Delta_{NG}$-exact: proportional to
$\Delta_{NG}\left( \zeta\bar\zeta\
\sum_{k=1}^\infty {\rm Re}\ h_k\zeta^{k-1}\right)$.
Solving equations (\ref{recuproc}) with the r.h.s. of this
form reduces to an exercise with the zero-modes, which
are explicitly known from (\ref{NGharf}).
The badly needed evaluation of higher order corrections to
(\ref{Aans}) can be more complicated, though many terms in
NG equations also become $\Delta_{NG}$-exact after
substitution of the previous-order solution.

A somewhat tricky part is integration, where it is instructive
to perform integration by parts and apply NG equations
whenever possible: in this way one can separate boundary
contributions, which should be thrown away in admissible
regularization schemes -- and eq.(\ref{Aans}) is written
under this assumption.
If regularization allows the boundary terms to contribute,
this causes a brutal violation of (\ref{AMD-BDS})
-- instead of a relatively "soft" violation in admissible
regularizations.

\section{Abelian Wilson average for a wavy circle}

Examination of BDS version of Alday-Maldacena duality also
requires evaluation of the double integral at the r.h.s.
of (\ref{AMD-BDS}).
The integral is also divergent and can be regularized,
say, by adding a term $\lambda^2$ to denominator
in the integrand: this $\lambda$-regularization is
a direct counterpart of the $\mu$-regularization of the area.
This is a considerably simpler calculation than that
of the area $A_\Pi$ at the l.h.s. of (\ref{AMD-BDS}),
it does not require solution
of any differential equations and is rather easily
extended to any particular order in $h_k$.
The first terms of this expansion are \cite{IMM2}:
\be
\begin{array}{ccccccccccccc}
{D_\Pi} & =& \frac{2\pi}{\lambda}L_\Pi &-& 4\pi^2 &
-&8\pi^2\Big( Q^{(2)}_\Pi &-&
Q^{(3)}_\Pi &+& Q^{(4)}_\Pi &+& O(h^5)\Big)
\label{Dans}
\end{array}
\ee where $L_\Pi$, $Q_\Pi^{(2)}$ and $Q_\Pi^{(3)}$ are the same
quantities that appeared in (\ref{Aans}). Thus we clearly see
the discrepancy between the two sides of (\ref{AMD-BDS}):
first, the coefficients in front of the brackets differ by a
factor of $\kappa_\circ = \frac{8\pi}{3}$, second, the diagonal
piece $Q_\Pi^{(3,diag)}$ of $Q_\Pi^{(3)}$ appears separately in
(\ref{Aans}), but not in (\ref{Dans}). This is a clear
violation of (\ref{AMD-BDS}): 
\be 
A_\Pi \neq D_\Pi, 
\ee 
even if
regularizations are matched, $\kappa_\circ\lambda = 4\mu$, and
unphysical constants $2\pi$ and $4\pi^2$ are omitted. However,
in some sense it is a {\it small} violation: $Q_\Pi^{(2)}$ and
$Q_\Pi^{(3)}$ are non-trivial non-local expressions, involving
infinitely many independent structures (different terms in
infinite sums) and only two of the infinitely many coefficients
in the so far evaluated orders  $h^2$ and $h^3$ distinguish
(\ref{Aans}) from (\ref{Dans}). One could even think that the
overall coefficient $\kappa_\circ$ is not a problem, but in
fact it is, if one believes that the relative coefficient
$\kappa$ between regularized exponents  at the two sides of
(\ref{AMD-BDS}) is independent of the shape of $\Pi$. The
problem is that in the $n=4$ example \cite{AM1,MMT1}
$\kappa_\Box = 8$ and it looks like $\kappa_\circ =
\frac{8\pi}{3} \neq\kappa_\Box = 8$ or 
$\frac{\kappa_\circ}{\kappa_\Box} = \frac{\pi}{3}\approx 1$.
Once again, the difference is at the level of $5\%$,
but it exists!

It would be very useful to reveal the general structure
of this "anomaly", what would hopefully allow to find
a relatively simple formula for the difference
between $A_\Pi$ and $D_\Pi$.
This requires calculation of at least some more terms
of the $h$-expansions at both sides.
Actually, some of the next corrections to $D_\Pi$ were evaluated in
\cite{IMM2} and this provides additional insight about the
structure of the formulas. For example,
$$
Q^{(4)}_\Pi = (h_1^2+\bar h_1^2)Q^{(2)}_\Pi
+ \frac{1}{4}\sum_{\stackrel{i,j,k,l=0}{i+j=k+l}}^\infty
U_{ij;kl}h_ih_j\bar h_k\bar h_l +
\frac{1}{6}\sum_{i,j,k=0}^\infty V_{ijk}
\Big(h_ih_jh_k\bar h_{i+j+k-2} + \bar h_i\bar h_j\bar h_k h_{i+j+k-2}
\Big),
$$ $$
U_{ij;kl} = \delta_{i+j,k+l}
\left(kC_{ij} - \frac{1}{6}(i+j)(k+1)k(k-1)(k-2)
+ \frac{1}{10}(k+2)(k+1)k(k-1)(k-2)\right),\ \ \ \
{\rm for}\ k\leq i,j,
$$
\vspace{-0.3cm}
\be
V_{ijk} = \frac{ijk}{3}\Big(i^2+j^2+k^2 +
3(ij+jk+ik) - 9(i+j+k) + 15\Big)
\ee
In particular, it is now clear that the split between
diagonal and off-diagonal parts, which happened to $Q^{(3)}_\Pi$
in (\ref{Aans}) but not in (\ref{Dans}), is not actually
a property of the area $A_\Pi$ alone: beginning from $Q^{(4)}_\Pi$,
it also occurs in expression for $D_\Pi$ in the Wilson average.
Once again, it would be very interesting to understand
the puzzling structure of these formulas, the meaning of these
splittings and of the coefficients $B, C, U, V, \ldots$
Possible insights about this structure can be provided
not only by calculation of the higher-order terms of
the $h$-expansion, but also by examination of parametrizations,
which are different from  (\ref{holpar}) and, further,
by consideration of non-planar deviations from a circle,
i.e. by wavy circles $\Pi$ which do not necessarily
lie in the circle's plane.

\section{Conformal symmetry}

Still another puzzle arises when we try to establish
conformal symmetry of the quantities $A_\Pi$ and $D_\Pi$.
On general grounds the both sides of (\ref{AMD-BDS})
are expected to possess this amusing symmetry \cite{DKS1}.
This was explicitly checked for $n=4$ in
\cite{Koma}, moreover the difference between abelian and
non-abelian Wilson averages for $n=6$, involved into
denunciation of BDS conjecture in \cite{B-V,DHKS03},
is also shown to possess this symmetry.
The more amusing becomes the problem, encountered in \cite{IMM2}
in attempt to test the naive version of this invariance.

If the boundary $\Pi$ is lying within a complex plane $z$,
conformal symmetry is induced by
the coordinate transformations $\delta z \sim z^p$ with $p=0,1,2$
and acts on the functions of $h_k$ by generators
\be\label{cs}
\hat J_{-}  = \frac{\partial}{\partial h_0}, \nn \\
\hat J_0 = \frac{\partial}{\partial h_1} + \sum_{k=0}^\infty h_k
\frac{\partial}{\partial h_k}, \nn \\
\hat J_{+} = \frac{\partial}{\partial h_2} + 2\sum_{k=0}^\infty
h_k \frac{\partial}{\partial h_{k+1}} +
\sum_{k,l=0}^\infty h_kh_l\frac{\partial}{\partial h_{k+l}}
\ee
Invariance of the function like (\ref{Dans}) requires that
the coefficients of $Q^{(m)}_\Pi$ are related:
$$
C_{1k} = B_k,\ \ \   C_{11}=C_{12} = 0, \ \ \
C_{2k} = 2B_{k+1},
$$ $$
V_{1kl} = 2C_{kl}, \ \ \
V_{2kl}+2A_{k+l}=2\left(C_{k,l+1}+C_{k+1,l}\right),
$$ \vspace{-0.6cm}
\be
U_{ij;1l}=C_{ij},\ \ \ U_{ij;2l} = 2C_{ij}
\label{coresym}
\ee
-- and this is indeed true for the actual values of
these coefficients.
Thus the only source of non-invariance in $D_\Pi$ is divergent
term, which is proportional to the non-invariant length
$L_\Pi = \oint_\Pi dl$,
transformed as
$$
\hat J_{-} L_\Pi = 0,
$$ $$
\hat J_{0} L_\Pi = \frac{1}{2}L_\Pi,
$$
\vspace{-0.4cm}
\be\label{sV}
\hat J_{+} L_\Pi = \oint_\Pi z dl = h_0L_\Pi + 2\pi\left(
(1+h_1)^2{\bar h_2\over |1+h_1|}+\sum_{k=2}{(k+1)(k+2)\over 4}
{1+h_1\over |1+h_1|}h_k\bar h_{k+1} + \ldots \right)
\ee
In fact, invariance properties of $D_\Pi$ are not quite
obvious from its double-integral representation in
(\ref{AMD-BDS}).
Invariance under the constant shift of $z = y_1+iy_2$ is evident,
while that under a dilation of $z$ is violated by
$\lambda$-regularization in an obvious way.
The most interesting is of course the action of the last
generator $\hat J^+$.
The point is that under the variation
$z \rightarrow z - \bar\beta z^2$
the integrand in the double integral changes by a total
derivative, for example,
\be
\oint\oint\frac{dzd\bar z'}{|z-z'|^2} \ \longrightarrow\
\oint\oint \frac{1-2\bar\beta z}{1-\bar\beta(z+z')
\phantom{.^{5^5}}\!\!\!\!\!\!}
\frac{dzd\bar z'}{|z-z'|^2} =
\oint\oint\frac{dzd\bar z'}{|z-z'|^2} - \bar\beta
\oint dz \oint\frac{d\bar z'}{\bar z-\bar z'}
+ O(\bar\beta^2)
\ee
This means that -- if not the divergency and associated
regularization -- $D_\Pi$ would be annihilated by $\hat J_+$
and that it finite (regularized) part actually is.

The puzzle arises when we switch to the minimal area
(\ref{Aans}).
Adding the new term with $Q^{(3,diag)}_\Pi$ to
$Q^{(3)}_\Pi$ can be considered as a change of coefficients
$C_{ij}$ for $C^A_{ij} = C_{ij} + 3C_{ii}\delta_{ij}$.
These modified $C^A$ should now appear in
(\ref{coresym}) instead of $C$,
and one of these relations is violated:
\be
C_{22}^A = 4C_{22} \neq 2B_3 \ \stackrel{(\ref{coresym})}{=}
\ C_{22}
\label{coresumvio}
\ee
Note, that the total rescaling of all
$B$ and $C$ by $3$ does not affect linear relations
(\ref{coresym}).
Thus (\ref{coresumvio}) is actually the only
violation of conformal symmetry which can be seen
at this level of accuracy (with $h^4$ terms neglected),
but it is enough to signal that there is a problem.
The problem is that there is a general belief
that it is the regularized area which should be
conformal invariant -- and we see that it is not.

The resolution of this puzzle is not yet known.
A possible clue can be once again related to regularization:
since the area integral diverges, it requires regularization,
which actually forces one to go away from the boundary
$\Pi$, i.e. away from the absolute of $AdS$, at least
a little.
However, away from the boundary  the action of conformal
group is different: instead of $\delta z \sim z^2$
we have \cite{Koma}
\be
r \rightarrow \frac{r}{1+2\vec\beta\vec y
+ \vec\beta^2(r^2+\vec y^{\,2})},
\nn \\
\vec y \rightarrow \frac{\vec y + \vec\beta(r^2+\vec y^{\,2})}
{1+2\vec\beta\vec y + \vec\beta^2(r^2+\vec y^{\,2})}
\ee
This means that outside the boundary $z = y_1+iy_2$
does not vary harmonically: there is additional
term $\sim r^2$ in the variation
$\delta z = -\bar\beta z^2$,
$\delta\bar z = +\bar\beta r^2$,
where non-holomorphic solution $r(z,\bar z)$ of NG
equations should be substituted.
Such non-harmonic $\delta \bar z$ is inconsistent with
parametrization (\ref{holpar}), and additional
coordinate transformation transformation is needed
to restore it.
To find this additional transformation is a separate
problem, because of the non-constructive nature of
Riemann theorem behind (\ref{holpar}).
All this implies that some $\mu$-dependent terms
can arise in the definition of operators $\hat J$
in (\ref{cs}), and they can contribute to the variation
of $A_\Pi$ because $A_\Pi$ contains a divergent term
$\sim \mu^{-1}$.
For {\it such} resolution of the symmetry problem to work
one should actually find a $\mu$-linear correction to
$\hat J$, $\hat J \rightarrow \hat J + \mu \hat j$,
-- which does not look like an obvious thing to exist,--
only then there could be a hope that $\hat j_+ L_\Pi$
compensates for the difference (\ref{coresumvio}).
Otherwise some {\it other} resolution of the controversy
between  (\ref{coresumvio}) and the beliefs behind
\cite{Koma} should be found.

The last comment about the symmetries of $A_\Pi$
and $D_\Pi$ is on a possible extension of finite-dimensional
$AdS$ conformal symmetry to something like the full-scale
Virasoro constraints (what is by now a conventional road
to take in the search of integrability structures
behind effective actions  \cite{gentau}).
From this perspective (\ref{cs}) seems to imply an obvious
possibility:
to make use of the Virasoro generators
\be
\hat J_m = \frac{\partial}{\partial h_{m+1}}
+ (m+1)\sum_{k=0}^\infty h_k\frac{\partial}{\partial h_{k+m}}
+ \ldots =
\sum_{p=0}^{m+1} \frac{(m+1)!}{p!(m+1-p)!}
\sum_{k_1,\ldots,k_p=0}^\infty
h_{k_1}\ldots h_{k_p}
\frac{\partial}{\partial h_{k_1+\ldots+k_p+m+1-p}}
\label{virext}
\ee
However, it is easy to see that the corresponding analogue
of (\ref{coresym}) is not true for $m>1$, already
\be
C_{m+1,k} \neq (m+1)B_{m+k}\ \ \ \ \ {\rm for}\ \ m\geq 2
\ee
Thus, if there is some Virasoro extension of (\ref{cs})
which can leave $A_\Pi$ and/or $D_\Pi$ intact,
it is not going to be as simple as (\ref{virext}).

\section{The hopes}

As already explained in s.\ref{presta}, the main problem with
Alday-Maldacena duality at the moment is the lack of its
more or less constructive {\it formulation}.
The ABDK/BDS conjecture \cite{ABDK,BDS}
about exponentiation of one-loop
amplitudes was absolutely distinguished, because it
allowed straightforward continuation of exact amplitudes
from the weak coupling regime, where they are given by
the sum of perturbative Yang-Mills diagrams,
to the strong coupling regime, where the dominant
contribution comes from the Gross-Mende \cite{GM}
minimal string surfaces.
The only undefined ingredient in that approach was a
{\it universal} redefinition of coupling constant, which
-- by universality -- could be borrowed from the deeply
investigated case of twist-two anomalous dimensions
\cite{anodi}, where generally expected integrable structure
\cite{gentau} is already well seen and intensively exploited.
However, today, after BDS conjecture is shown to be wrong for
$n\geq 6$ \cite{B-V},
-- as anticipated in \cite{DKS1} and
pointed as the only way out of emerging contradictions
in \cite{AM3,DHKS01,DHKS03,IMM2},--
the situation is far more complicated.
Even if for some yet-unknown reason all the multi-loop
diagrams in perturbation theory do
sum up into non-abelian Wilson average
-- as predicted in \cite{DKS1,DHKS01,B-V,DHKS03},--
instead of the abelian one
-- as too optimistically conjectured in \cite{BDS}, --
non-abelian average is far more difficult to extrapolate
to the strong-coupling phase:
even plausible conjectures about result of such continuation
are still absent.
This means that there is no constructive formula at the r.h.s.
of the modified Alday-Maldacena duality relation
(\ref{AMD-DHKS}) to compare with the regularized minimal area
at its l.h.s.

So, what can we do -- except for abandoning the very
story of Alday-Maldacena duality
until better times, when new ideas emerge, or forever, if they
do not?

An obvious possibility is to look under the lamp:
one can carefully study the two sides of the (failed)
relation (\ref{AMD-BDS}) and try to rewrite the difference
between them as some correction to the double-loop
integral at the r.h.s., perhaps, as a combination
of correction to the {\it integrand} and contribution of
additional integrations, or in some other way.
In case of success we can probably understand what is
actually standing at the r.h.s. of duality relation,
i.e. what is the possible (boundary integral?) reformulation
of solution to Plateau problem.
If such reformulation exists, one can switch to
the reason of its relation to non-abelian Wilson average
at weak coupling.
It goes without saying that existence of such reformulation
is far from obvious from the perspective of Plateau problem
and by itself it would be a great achievement in the theory
-- and of course the failure would not be a big surprise.
Still, the very spirit of string-gauge duality provides
a hope: some relation like Alday-Maldacena duality should
exist and previous experience implies that in the case
of ${\cal N}=4$ supersymmetry it should be a reasonably
simple relation, perhaps, not as simple as (\ref{AMD-BDS}),
perhaps even with complexity comparable to
the theory of coupling renormalization \cite{anodi},
but hopefully not beyond it.
Somehow, the results of the first such calculation,
reported in \cite{IMM2} and above, do confirm this expectation:
the difference between the violation of BDS version
of Alday-Maldacena duality (\ref{AMD-BDS}) is essential,
but it corrects only a few structures in the
infinitely-parametric family, i.e. is much weaker than it
could be.
Moreover, corrected structures are not distinguished by
no one of the two known tiny (conformal) symmetries and
looking at these structures one can probably find out what
the right hidden symmetry is.
However, in order to go this way, one needs an
independent validation of rather lengthy calculations in
\cite{IMM2} and their straightforward, still tedious,
extension to a few higher orders in $h$-expansion
(to get a broader view on the structures involved),
to non-planar deformations of a circle and also
to other examples,
in the spirit of \cite{IMM1} and \cite{IM8}.

\section*{Acknowledgements}

\noindent

I am indebted for hospitality to the organizers of the
workshop and to all its participants for a wonderful
and stimulating atmosphere of the meeting.
I acknowledge the support of JSPS and OCU during my
stay in Osaka.

This presentation reports the results of our common work
with Hiroshi Itoyama, Andrei Mironov and Theodore Tomaras,
to whom I am grateful for inspiration and other pleasures
of doing science together.

I acknowledge interesting discussions on the topics of
this text with many of my colleagues and friends, especially
with Katsushi Ito, Antal Jevicky, Alexander Gorsky,
Hikaru Kawai, Horatiu Nastase,
Arkady Tseytlin, Anton Zabrodin and Konstantin Zarembo.

My work  is partly supported by Russian Federal Nuclear Energy Agency
and Russian Academy of Sciences,
by the joint grant 06-01-92059-CE,  by NWO project 047.011.2004.026,
by INTAS grant 05-1000008-7865, by ANR-05-BLAN-0029-01 project,
by RFBR grant 07-02-00645 and
by the Russian President's Grant of Support for the Scientific
Schools NSh-3035.2008.2.

\end{document}